\documentclass[12pt]{svproc}
\usepackage{url}

\usepackage{natbib}

\bibpunct{(}{)}{;}{a}{}{,}%
\usepackage{graphicx}
\usepackage{amsmath,amssymb}
\usepackage{multirow}
\usepackage{longtable}

\usepackage{epsfig}
\usepackage{setspace} 

\usepackage[b5paper]{geometry}
\newcommand\blfootnote[1]{%
  \begingroup
  \renewcommand\thefootnote{}\footnote{#1}%
  \addtocounter{footnote}{-1}%
  \endgroup
}

\begin{document}

	\mainmatter  
	\title{Cosmological Model Tests with JWST}
	\titlerunning{Cosmological Model Tests}  
	\author{N. Lovyagin$^{1}$, A. Raikov $^{2}$, V. Yershov $^{3}$ and Yu. Lovyagin $^{1,4}$}
	\authorrunning{Lovyagin et al.} 
	\tocauthor{N. Lovyagin, A. Raikov, V. Yershov and Yu. Lovyagin}

     \institute{ 
        {\scriptsize Dept. of Computer Science, St. Petersburg State University, 7/9 
 Universitetskaya nab., 199034 St. Petersburg, Russia~~~~~~~~~~}
		\and
		{\scriptsize St. Petersburg Branch of Special Astrophys. Observ., Russian Acad. Sci., 65 Pulkovskoye sh., St. Petersburg 196140, Russia~~}
		\and
		{\scriptsize Moniteye UK, 30a Upper High Str., Thame OX9 3EX, United Kingdom~~~~~~~~~~~~~~~~~~~~~~~~~~~~~~~~~~~~~~~~~~~~~~~~~~~~~~~~~~~~~~~~~~~~~~~} 
	    \and
	    {\scriptsize Dept. of Mathematics, State Marine Technical University, Lotsmanskaya Str., 3, 190121 St.~Petersburg, Russia}~~~~~~~~~~~~~~~~~~~}	
	
	\maketitle

\begin{abstract}
\vspace{-1cm}
The James Webb Space Telescope (JWST), which has recently become operational, is capable of detecting objects at record-breaking redshifts, $z \gtrsim 15$. This is a crucial advance for observational cosmology, as at these redshifts the differences between alternative cosmological models manifest themselves in the most obvious way. In recent years, some observational hints have emerged indicating that the Standard Cosmological Model could require correcting. One of these hints is related to the discovery of remote galaxies whose redshifts correspond to the very young Universe (less than one billion years after the Big Bang) but which are similar to nearby galaxies. The issue is that such galaxies in the early Universe do not have enough time to evolve into something similar to the late-Universe galaxies. JWST observations of high-redshift objects are expected to shed light on the origin of this issue. Here we provide results on performing the ``angular diameter---redshift'' cosmological test for the first JWST observation data. We compare this result with predictions of the standard $\Lambda$CDM cosmological model and some static cosmological models, including Zwicky's ``tired-light'' model. The latter is currently assumed to be ruled out by observations. We challenge this assumption and show that a static model can provide a natural and straightforward way of solving the puzzle of the well-evolved galaxies and better agreements with the results of the JWST ``angular diameter---redshift'' test at higher redshifts than the correcting evolution model within the $\Lambda$CDM framework. 
We discuss several cosmological tests that will be important for further research on the possibility of revising 
 the expanding Universe paradigm.
	\keywords{James Webb Space Telescope; observational cosmology; galaxies; standard cosmological model; tired-light model; redshift--distance relationship; surface-brightness test; angular-diameter distance}
\end{abstract}

\section{Introduction}

One of the main scientific goals of the James Webb Space Telescope (JWST)  is to
explore the Universe's history following the end of the period of the so-called ``dark ages''. 
\blfootnote{{ Galaxies} {\bf 2022}, 10(6), 108 \texttt{~~https://doi.org/10.3390/galaxies10060108}} 
\setcounter{footnote}{0}
JWST has been designed to detect the appearance of the first sources of light in the Universe  and  uncover the history of the assembly of first galaxies \citep{gardner06}.     
An analysis of observations made by large telescopes prior to JWST  indicated that the first stars and galaxies might have appeared 
between 250 and 350 million years ($z>15$) after the Big Bang \citep{laporte21,ellis22}.
Indeed, based on the first JWST observations, Donnan et al. \cite{donnan22} 
report a newly discovered galaxy with a redshift $z \sim 16.7$. This redshift
corresponds to approximately 250 million years after the beginning of the Universe.
In principle, JWST was designed to detect objects in the early Universe, 
only several tens of million of years old (should such objects exist). This is achievable 
because  the red wavelength cut-off of the NIRCam instrument onboard JWST extends to 5\,$\mu$m.

Another group of researches, \cite{atek22}, has reported the discovery of two
galaxy candidates at $z \sim 16$, two candidates at $z = 12$ and eleven candidates 
at $10 < z < 11$ (these redshifs have been estimated photometrically for the 
galaxies gravitationally lensed by the massive galaxy cluster SMACS\,J0723-73). 
The morphologies of these 
high-redshift galaxies turn out to be consistent with disks, while their sizes 
are smaller compared to similar galaxies at lower redshifts. 
The unexpected excess of bulges and disk-shaped galaxies at high redshifts has also been 
confirmed by the morphological study by \cite{jacobs22} of 217 sources at redshifts 
$1 < z < 5$. 

A flanking field around the same cluster SMACS\,J0723-73 (not magnified by gravitational lensing) 
has been studied by  \cite{yan22}, who searched for galaxies at a redshift
larger than 11 and which found 88 candidates, some of them might be at redshifts 
as high as $z=20$. The large number of such objects at high redshifts was not 
previously predicted by standard cosmology. 

By contrast,  \cite{castellano22} found a smaller number of high-redshift
galaxies on a flanking field around the Frontier Fields cluster A2744: nine objects at $9 < z < 12.3$,
two of the brightest of them at $z > 10$, being unexpected given the survey volume.

Spectroscopic studies of three remote gravitationally lensed galaxies at $z=7.7$ and $z=8.7$ 
within the field of the cluster SMACS\,J0723-73 \citep{shaerer22}
reveal a strong resemblance of emission line properties 
to the spectra of their local-Universe counterparts.
Similarly, the measurements of the rest-frame ultraviolet continuum slopes 
of galaxies at $8 < z <15$ show that these galaxies are no bluer than the bluest 
galaxies in the local Universe \cite{cullen22}. These slopes are indicators 
of ultra-young stellar populations, which are expected to be prevalent in the high-redshift 
Universe---but they are not. 

Measurements of remote galaxy masses and sizes (half-mass radii) suggest
an inverse relationship between these quantities; that is, the most massive
high-redshift galaxies are more compact and dense  \citep{marshall22}.
This study has been made for galaxies at redshifts $z=7$ to 11 prior the first 
JWST data release. A similar study based on the JWST data 
\citep{suess22} also reveals that the high-redshift galaxies are very compact and 
massive, showing the same trend (i.e., smaller galaxies having larger masses). 
These authors conclude that their result impacts our understanding of the size growth and evolution 
of galaxies in the early Universe.    

It is noticeable that most of  these studies of the first JWST data release have something in common.
Namely,
\begin{itemize}
\item
There is an excessively large number of galaxies at very high redshifts, which is not foreseen by 
the Standard Cosmological Model;
\item
Galaxies at these redshifts have disks and bulges, which indicates that they have passed through a long 
period of evolution;
\item
Spectroscopically, these galaxies resemble their counterparts in the local Universe;
\item
Smaller galaxies are more massive than larger ones, which is quite the opposite of the common view.   
\end{itemize}

These issues indicate that the galaxies at redshifts $z>15$ discovered by JWST do not have enough time within the framework of the standard cosmological
model to evolve into what is observed.

Even before the operational period of JWST, other large telescopes, such as HST (the 
Hubble Space Telescope) or VLT (the Very Large Telescope),  were finding an
ever-increasing number of high-redshift objects, whose age from the beginning of the Universe was below a billion years, and  
whose formation within the $\Lambda$CDM cosmological model is difficult to explain
\citep{disney12,shibuya15,andreon18}. 
These objects are fully evolved, very large and very bright  galaxies,
ultra-luminous optical and X-ray quasars with the masses of their central 
blackholes reaching a few billion solar masses.  

The existence of such objects would require new models of their 
formation. Alternatively, some brightness-amplification effects,
such as gravitational lensing of high-redshift quasars \citep{raikov21},
might reduce the estimated 
masses of supermassive blackhole populating the early Universe. However, this effect
cannot help explain complicated morphologies of high-redshift galaxies. 
Another alternative for the explanation of supermassive blackholes 
in the early Universe is to postulate the appearance of primordial blackholes (PBH)
at the very beginning of the Universe's existence, before the time of recombination 
or even before the beginning of baryonic acoustic oscillations (BAO),
which are currently regarded as the main cause of structure formation 
in the early-Universe.  
The main objective of the PBH model promoted by \cite{dolgov18} 
is to explain the existence of well-evolved objects whose formation 
is believed to be impossible within the $\Lambda$CDM framework 
due to the very short period of available time \footnote{The same work also provides a 
detailed overview of theoretical constrains on structure formation time due to BAO 
within the $\Lambda$CDM framework.}.

This lack of evolution time is also a problem for known quasars containing supermassive blackholes 
with masses exceeding $10^{10}$\,M$_\odot$ at redshifts $z>6$ \citep{wu15,banados18,yang20}, 
whose existence is inconsistent with their age of shorter than a billion years after the Big Bang.
This is also a  challenge to the standard cosmological model
itself \citep{dolgov18,dolgov20}.  In principle, the issues of very small, but very massive and 
well-developed galaxies seen at very high redshifts, could be solved by ad hoc adjustments 
to galaxy formation and growth models. However, a much simpler, although quite radical solution,
might be found by shifting the paradigm from an expanding to static Universe, as was proposed
by  \cite{laviolette21} and some other researches 
\cite{crawford14,lopezcorredoira17,lerner18,lopezcorredoira22}, 
including one of the authors of the present article \citep{orlov16}.

The JWST is expected to detect light emitted by the first stars in the Universe,
when the first galaxies or protogalaxies were coming into existence. 
This prediction is based on the standard $\Lambda$CDM cosmology. However, the existence of well-developed galaxies, should they be detected by JWST, is not foreseen within the framework of 
$\Lambda$CDM.

Here we shall analyse the possibilities provided by the JWST for testing cosmological models
using ultra-high redshift objects and comparing the observed 
photometric, spectrophotometric and geometric parameters of these objects with the predictions
of the standard Lambda--Cold--Dark Matter model ($\Lambda$CDM) 
and some alternative cosmological models.   Throughout this paper,
we use a standard cosmology with the parameters $H_0=70$\,km\,s$^{-1}$\,Mpc$^{-1}$
(the Hubble constant), $\Omega_\Lambda =0.7$ (the dimensionless density of dark energy) and 
$\Omega_{\mathrm M} = 1 - \Omega_\Lambda =0.3$ (the density of matter, including both 
baryonic and dark matter), assuming a flat Universe with the curvature energy density
$\Omega_k=0$.

\section{Materials and Methods}

\subsection{Observational Data for the Early-Universe Objects}
\label{sec2.1.}
The observations used for our analysis are publicly available  JWST datasets, which 
include NIRCam images in F090W, F150W, 
F200W, F277W, F356W and F444W filters; MIRI images in F770W, F1000W, F1500W and F1800W filters; 
NIRSpec spectra in F170LP and F190LP, as well as NIRISS spectra obtained with F115W 
and F200W filters. These data were released on 12th of July, 2022 
at \url{https://webbtelescope.org/contents/news-releases/2022/news-2022-035} (accessed on 30 July 2022), 
as well as at the
{\tt Mikulski Archive for Space Telescopes} (MAST) \footnote{\url{https://archive.stsci.edu}, accessed on 1 October 2022.} 
under program ID 2736.
The associated programmatic interface 
 \footnote{\url{https://astroquery.readthedocs.io/en/latest/mast/mast.html}, accessed on 1 October 2022.}
provides scripts for the data access and reduction. 
Some results of the JWST data reduction are also made publicly available. For example,
the calibrated and distortion-corrected NIRCam and NIRISS images processed by G. Brammer
are accessible  \url{https://s3.amazonaws.com/grizli-v2/SMACS0723/Test/image_index.html} (accessed on 1 October 2022).  
The catalogues of ultra-high-redshift objects detected by the JWST instruments in the 
SMACS-0723 deep field are also publicly available at \url{https://zenodo.org/record/6874301\#.YubQUfHMJes} (accessed on 1 October 2022). 

We make use of the preliminary results from these data analyses  
published by various research groups, mainly in the form of arXiv e-print manuscripts 
at \url{https://arXiv.org} (accessed on 1 October 2022). Most of these authors report an unexpectedly large number 
of well-evolved galaxies at redshifts corresponding to their age from the beginning of the Universe of 
$\sim$200--250 Myrs. 
JWST images and spectra provide information about photometric and geometric parameters 
of remote galaxies, such as their brightnesses, sizes and redshifts,  which  have 
already been estimated and published by other authors.     
The accuracy of available photometric redshifts is not very high. Most of the redshift error bars on photometric 
redshifts in the recent publications on JWST data are within the range $0.1 < \sigma_z <2.0$, 
while spectroscopic redshifts  estimated by using 
spectral lines identified from the JWST/NIRSpec data have $\sigma_z< 0.01$, see, e.g., \cite{tacchella22}.    
However, some of the redshifts are  unreliable, with $\sigma_z > 6$ 
(see  { Section 3}).
For example, one of the JPWS high-redshift galaxies can be fit with the redshift either 
$z \approx 17$ or $z \approx 5$
\citep{naidu22a}.
In our analysis here, we use approximate redshifts from the published JWST galaxy data with  
$\sigma_z < 3$.

\subsection{Cosmographic Tests}

The methodology underlying our analysis is based on   { cosmographic} 
theories
described in classical \citep{tolman34,mcvittie56,zeldovich67, harwit73, zeldovich75, peebles93}  and modern 
\citep{raine01, narlikar06, baryshev12,gabrieli05} textbooks, as well as in some dedicated 
reviews { \citep{lopezcorredoira16}} and papers cited below. Here we shall focus primarily 
on the   { angular size--redshift} relationship, $\theta(z)$.

This relationship has been widely used for comparing different cosmological models.
For example,  \cite{nabokov08} measured angular sizes of 
galaxies in the Hubble ultra deep field for $ 0.5 < z < 6.5$ with the purpose to 
find inhomogeneitis  in their radial distribution  and concluded  that the the current 
model of the evolution of galaxy sizes is not yet reliable enough for using $\theta(z)$
as a cosmological test at the studied redshifts. However, they noted that at $z \approx 6.5$
the measured galaxy angular sizes do not match well with the predictions of the standard cosmological model. 
 \cite{lopezcorredoira10} also noted that there is degeneracy between expansion 
combined with galaxy size evolution and non-expansion. Furthermore, he showed that  a simple
static model with no evolution in size and no dark matter ratio variation fits  the 
observed $\theta(z)$ relationship better than the standard model. 

Other tests, such as the Tolman  surface-brightness test, the cosmological time dilation; 
 number density--redshift relationship, 
galaxy-number-count--magnitude; photon-flight-time--redshift relationship
and some others are of equal importance. 
 We shall postpone their discussion to future works, as the angular-size test alone 
already provides insight into the problem. 

The main purpose of these tests is to shed light on the origin of the cosmological redshift.
This could be due to the growth with time of the global scale factor of the Universe, 
which can be viewed in the form of radial velocities of all galaxies
with respect to each other. Alternatively, the cosmological redshift might be caused 
by some physical effects, such as possible photon energy dissipation along the photon's path 
or photon energy
change in gravitational potential wells. 

Accordingly, cosmological models can be divided in two groups: 
\begin{enumerate}
\item
Expanding universes based on the Friedmann–Lemaitre–Robertson–Walker (FLRW) metric
with a time-dependent scale factor;
\item
static universes based, e.g., on the metric including a scale factor in metric's time component 
\citep{troitskij95}
or Zwicky's model based on the photon-energy dissipation along the photon's travelling path \citep{zwicky29}.  
\end{enumerate}
A mixture of these two model types is also possible { \citep{gupta18a,gupta18b} } when a physical effect enhances the
redshift due to the growing scale factor. In which case, the expansion rate of the Universe would 
be smaller than predicted by the observed cosmological redshift within the 
expanding-Universe model. 
Correspondingly, the age of the objects in the Universe in a mixed-type model could be 
larger than the age deduced from pure FLRW 
models, which would mitigate the problem of the well-evolved galaxies discovered by the JWST
at ultra-high redshifts, not having enough  time for their formation and evolution.   

The commonly accepted model of the first type is the
standard $\Lambda$CDM cosmological model, which best fits observational data 
among other expanding-Universe models \footnote{nevertheless, we shall see that it fails 
to fit the recent JWST observations.}. Those other models played an important role in the past 
for the development of the methods of observational cosmology. Therefore, 
perhaps it is worthwhile mentioning one of them---the  { steady-state} 
cosmological model,
first proposed by A. Einstein  in 1931 \citep{einstein14}\footnote{although abandoned by him in favour 
of his other expanding-Universe model \cite{einstein31}.} and then in 1948, independently
by \cite{bondi48} and  \cite{hoyle48}.  
In this model, the Universe is stationary (although expanding) at the expense of
the proposed continuous creation of matter. 
It was well-elaborated in detail, but was failing to pass through 
cosmological tests. Thus, it required introducing numerous
additional features \citep{hoyle92,hoyle93,hoyle94a,hoyle94b,hoyle95}
and eventually was abandoned.

The most discussed model of the second type is Zwicky's model based on the idea 
of a photon's energy dissipation, which is not
commonly accepted, but which was found to be the best fit of all cosmological tests
together \citep{laviolette21} when compared with the same tests applied to the $\Lambda$CDM
model. 

Here we shall discuss the two possibilities---expanding or static Universe---
because the predictions of cosmological model tests for them are known 
to be very distinct at the high redshifts achievable by the JWST.
The most obvious distinction consists of the predicted observed angular size
of a galaxy as a function of its redshift 
\citep{devaucouleurs48,zwicky57,hoyle59, hickson77,kapahi87}, both angular size 
and redshift being directly observable quantities.

This difference in the predicted angular sizes of galaxies obviously affects their  
surface brightnesses. Therefore, the Tolman surface brightness test 
\citep{tolman30, hubble35} would be among other
important tests for distinguishing the origin of the cosmological redshift.    

Another significant distinction between the expanding- and static-Universe models 
consists of the relationship between the increase in the line-of-sight distances  
corresponding to redshift increments $\Delta z$. In the FLRW models, 
the physical-to-comoving volume ratio is strongly reduced 
at high redshifts. Since, by definition, the number of objects
locked in a comoving volume is constant, the number-density of these objects in the 
corresponding physical volume would be dramatically increasing in expanding-Universe 
models. Thus, the 
number-count of high-redshift galaxies observed by the JWST can serve as yet another 
cosmological test for distinguishing between the two redshift types.    
Although this kind of test is complicated by the fact that the galaxy number 
densities at high redshifts are related to the galaxy formation through the 
number of galaxy systems formed during their evolution, the ultra-high 
redshifts accessible by the JWST can help disentangle the evolutionary- and 
cosmological-model-related issues.

\subsubsection{ Angular Diameter---Redshift Relationship in the $\Lambda$CDM Model.}

The observable cosmological-distance measure is the cosmological redshift $z$
defined as the difference between the wavelength $\lambda_{\tt obs}$ measured in the coordinate
frame of the observer and the wavelength $\lambda_{\tt src}$   
emitted by a remote source:      
\begin{equation}
z= \frac{\lambda_{\tt obs} - \lambda_{\tt src}}{\lambda_{\tt src}}\,.
\label{eq:redshift_definition}
\end{equation} 
\noindent
The source is assumed to be at rest with respect to the Hubble flow---the coordinate frame moving away from the observer with the recession speed \citep{davis01}
\begin{equation}
v = \frac{c}{a_0} \frac{d a(t)}{dt} \int\limits_0^z \frac{dz'}{H(z')}\,,
\label{eq:recession_speed}
\end{equation}
\noindent
which is $\approx H_0 D$ for small $z$, where $D$ is distance (in Mpc) in the expanding Universe;
$a(t)/a_0$ is the normalised scale factor; and $H(z')=(1 + \Omega_M \left[ (1+z')^3 - 1 \right])^{-1/2}$.

The angular diameter distance $D_A$ of an astronomical object (from an observer)  is 
the ratio of the (perpendicular to the line-of-sight) 
physical linear size $\delta$ of the object  (e.g., its diameter) to its angular 
size $\theta(z)$ as measured by the observer: 
\begin{equation}
D_A(z)= \frac{\delta}{\theta(z)}\,,
\label{eq:angular_diam_dist}
\end{equation} 
\noindent
$z$ being the redshift of photons emitted from the object.
$D_A$ is a model-dependent quantity determined in the 
simplest approximation of the standard cosmology for a flat universe as
\begin{equation}
D_A(z)^{\Lambda{\rm CDM}}=\frac{c}{H_0} \frac{1}{1+z} \int\limits_0^z \frac{\mathrm dz'}{\sqrt{1 + 
\Omega_{\mathrm M} \left[ (1+z')^3 - 1 \right]}}.
\label{eq:angular_diam_integr}
\end{equation}  

\begin{figure}[h]
%
\hspace{1cm}
\includegraphics[width=8cm]{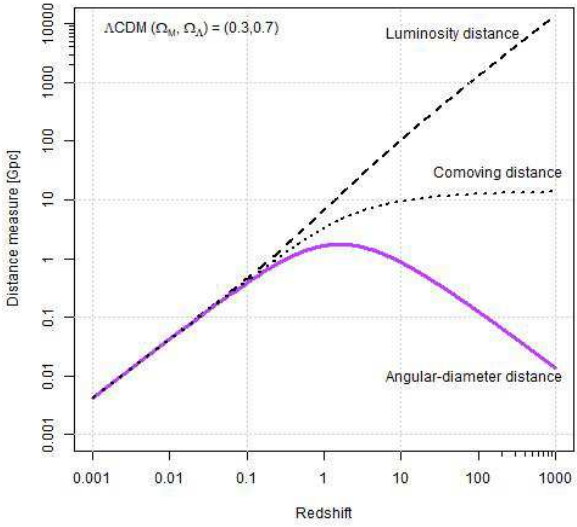} 
\caption{ Angular-diameter distance $D_A$ (purple curve) as calculated
within the $\Lambda$CDM model for $H_0=70$\,km\,s$^{-1}$\,Mpc$^{-1}$.
The luminosity distance $D_L$ (dashed line) and comoving distance $D_C$ (dotted curve) 
are also shown for comparison.
}
\label{fig_angdist}
\end{figure}

The angular diameter distance, as calculated for the standard $\Lambda$CDM model, is plotted 
in Figure \ref{fig_angdist} (the purple curve).
 At redshifts higher than $z \approx 1.61$,
the angular diameter distance diminishes  because  
the scale factor (``size'' of the Universe) is smaller at the moment of time when
light from a remote source is emitted than at the moment of time when this light is 
detected by the observer. 
\vspace{-10pt}
Figure \ref{fig_angdist} also shows two other cosmological distance measures:
\begin{equation}
\hspace{-6.2cm}\textrm{the comoving distance~~~~~~~~~~~~~~~~~} D_C(z) = (1+z) D_A(z)
\label{eq:dist_comoving}
\end{equation}
\noindent 
and the luminosity distance 
\begin{equation}
D_L(z) = (1+z)^2 D_A(z)\,,
\label{eq:dist_lumin}
\end{equation}
\noindent
the latter being defined as the relationship between the bolometric flux $F$
and the bolometric luminosity $L$:
\begin{equation}
D_L(z) = \sqrt{\frac{L}{4\pi F(z)}}\,. 
\label{eq:dist_lumin_flux}
\end{equation}

The expression (\ref{eq:dist_lumin}) is sometimes  called the Etherington's distance-duality relationship because
it is based on the reciprocity theorem for null geodesics proven by  \cite{etherington33}. 
It was explicitly identified by  \cite{mcvittie56} but was implicit in R.C. Tolman's 1930s works \citep{tolman30,tolman34}.

\begin{figure}[h]
%
\hspace{1cm}
\includegraphics[width=7cm]{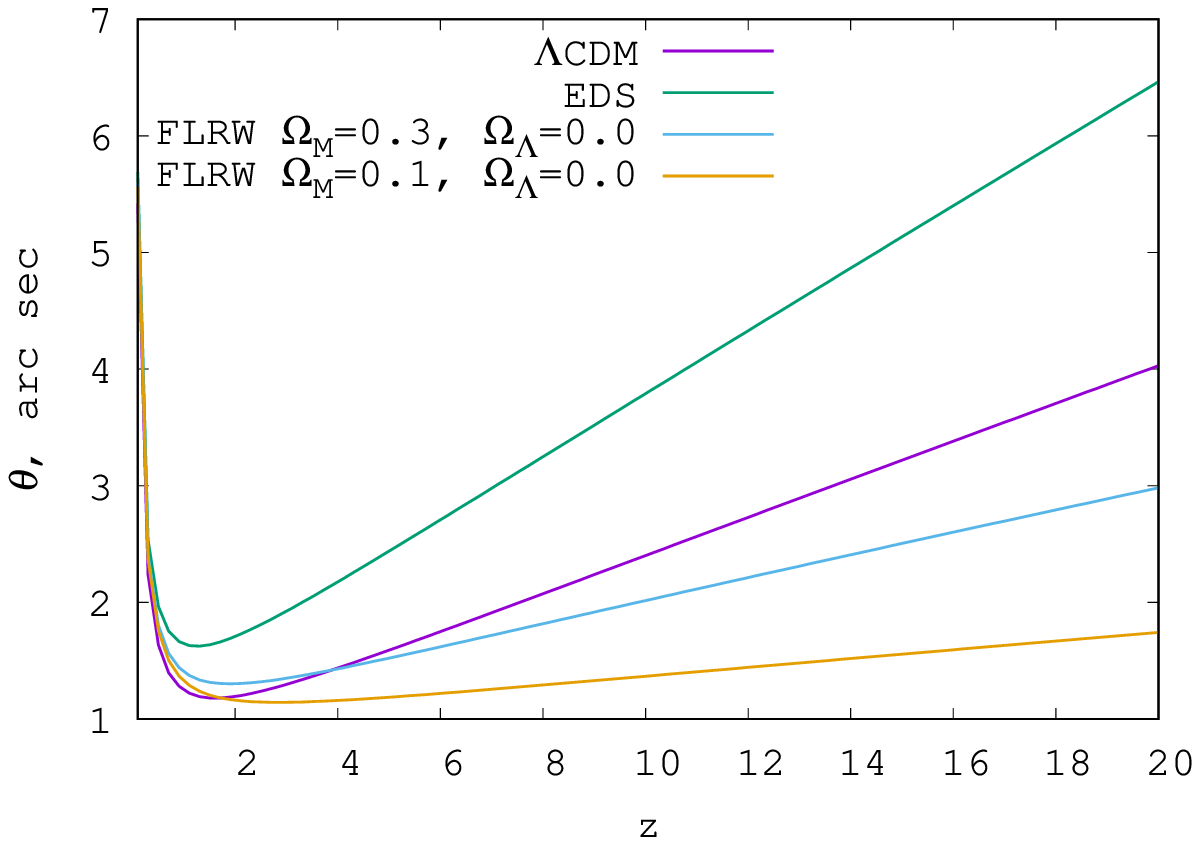} 
%
\hspace{-0.2cm}
\includegraphics[width=7cm]{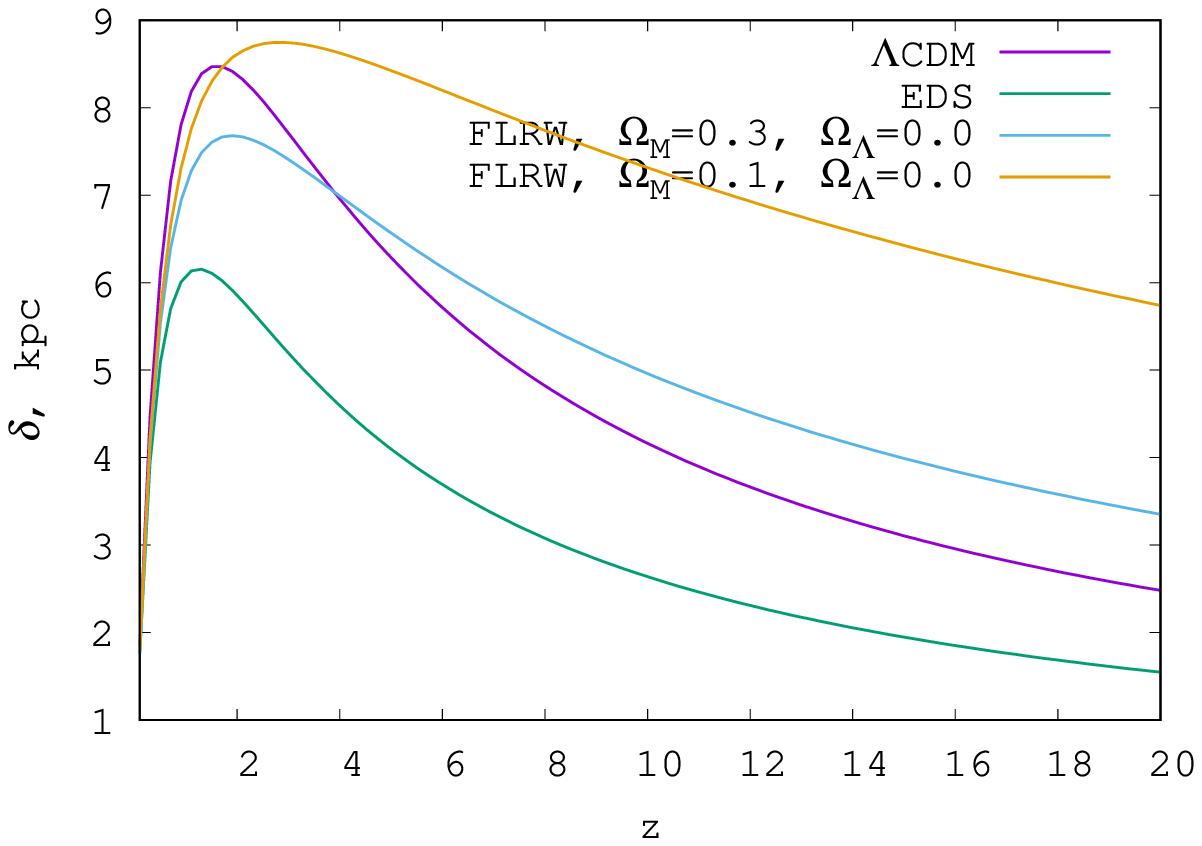}
 %
%
\vspace{0.6cm}
\caption{ Angular diameter of a 10 kpc-size object 
(left) and the linear diameter $\delta$ of a one-arcsec-size object (right) as functions of 
redshift $z$, corresponding to various models within the expanding-universe framework (FLRW).
The purple curves show the $\theta(z)$ relation for the standard $\Lambda$CDM model 
with $\Omega_{\mathrm M}=0.3$
and $H_0=70.0$ km\,s$^{-1}$\,Mpc$^{-1}$. The green curves correspond to the 
Einstein-de Sitter Universe ($\Omega_{\mathrm M}, \Omega_\Lambda)= (1,0)$. The blue and yellow
curves show $\theta(z)$ for two values of $\Omega_{\mathrm M}$ for FLRW models without dark energy.}
\label{fig1ab}
\end{figure}

The angular-diameter distance (\ref{eq:angular_diam_dist}) and angular size
$\theta$ are inversely related to  each other.
Therefore, the theoretical angular diameter 

\begin{equation}
\theta(z) = \frac{\delta}{D_A(z)^{\Lambda{\rm CDM}}}
\label{eq:dist_angdiam}
\end{equation}
\noindent
in the framework of the 
$\Lambda$CDM model is expected to be increasing at $z > 1.61$ for an astronomical
object of a fixed linear diameter $\delta$. 
This is illustrated by the plot of $\theta(z)$---the purple curve   
on the left panel of Figure \ref{fig1ab} for an object having a fixed size 
$\delta=10$\,kpc (slightly smaller than the Milky Way size).

Since it is the angular size $\theta$ that is the typical observable  
for remote galaxies, we can determine the linear size of an object with redshift $z$ 
by using the formula 
\begin{equation}
\delta(z) = \theta D_A^{\Lambda{\rm CDM}}(z).
\end{equation}
\noindent
The corresponding plot is presented in the right panel of Figure \ref{fig1ab}. It 
shows the linear measure (in kpc) for each arcsec of the apparent angular size 
of an object whose cosmological redshift is $z$. We can see that since 
the angular resolution of 
JWST is $\sim 0.1''$ \footnote{\url{https://www.jwst.nasa.gov/content/about/faqs/faq.html\#sharp}, accessed on 1 October 2022.},
this telescope can easily resolve the sizes of high-redshift galaxies for any FLRW model.

The $\Lambda$CDM framework suggests that the
JWST must find large images of remote galaxies  whose surface brightness
is low. However, what is currently observed is something opposite to what is expected:
there are small (by their angular size) galaxies with bright surfaces
at ultra-high redshifts.  
Perhaps this can be explained by modifying 
galaxy formation and evolution theories. However, for the sake of impartiality and simplicity,
one also has to check the congruency of JWST data to the static-
or slowly-expanding universe models with alternative physical mechanisms
of the cosmological redshift.

\subsubsection{Static-Universe Models.}

\noindent
In order to compare the $\Lambda$CDM-interpretation
of JWST observations with interpretations based on alternative cosmologies, we shall examine  
the most widely discussed alternative cosmology based on F. Zwicky's proposal 
in which he attributes the cosmological redshift to one of the possible 
physical mechanisms of photon-energy loss \citep{zwicky29}.

In his work, Zwicky analysed three possible physical mechanisms that could provide
the necessary energy loss of photons on their path through spacetime: 
\begin{enumerate}
\item 
Compton scattering on free electrons;
\item
Gravitational redshift due to gravitational
potential wells of galaxies or galaxy clusters along the photon's path;
\item
General-relativistic transfer 
of photon energy/mass to the masses distributed along the photon's path.
\end{enumerate} 

When checking the viability 
of these three mechanisms, F. Zwicky demonstrated that the first two of them were not helpful
in the explanation of the cosmological redshift and must be rejected.  Whereas the third 
possibility could still be regarded as a viable alternative to the FLRW-mechanism 
of the photons stretching in the expanding space. 
Later R. Tolman coined Zwicky's proposal as the ``tired-light'' (TL) theory.
Nowadays, the prevailing modern interpretation of this theory is based 
on the photon-scattering mechanism. 
Following this tradition, here we shall use the notation TL referring to
this particular mechanism of the cosmological redshft. It is commonly  believed
that the photon-scattering mechanism is likely to be wrong. It was rejected
straightaway by Zwicky himself in his original paper. One of the main
observations against this mechanism is the absence of blurring
of remote galaxy images. Indeed, photon scattering, e.g., on electrons, 
results in a significant photon-scattering angle. 
However, in principle, it is possible to explain the photon-electron 
interaction with the corresponding photon-energy loss without any scattering 
angle. For example, in Ashmore's theory \cite{ashmore15}, photons are absorbed and re-emitted 
by electrons in the intergalactic medium. The electron recoils and the photon loses
energy. There is no angular spread in this mechanism, as it is equivalent to
photon transmission in a transparent medium. According to \cite{ashmore15},
an electron density $\rho_0=0.8 h_{100}$\,[m$^{-3}$] produces the required 
cosmological redshift magnitude.         
Therefore, we assume here that the TL-model cannot be completely disregarded   
on the basis that it disagrees with observations.
This model is still under discussion, and its theoretical aspects are rigorously 
formulated by \cite{laviolette86}.
This theoretical consideration  leads to the following 
expression for the redshift-distance relationship:
\begin{equation}
z(r) =\frac{\Delta \lambda}{\lambda_0}= \exp({\beta r}) -1,
\label{eq:tl_redshift_dist}
\end{equation}
where $z$ is the redshift of the photon's initial wavelength $\lambda_0$
after the photon has travelled a distance $r$; and $\beta=H_0/c$ is the energy attenuation 
coefficient. The form of (\ref{eq:tl_redshift_dist}) is exponential 
because  photon-scattering  is cumulative  along the photon's path.  

Unlike the $\Lambda$CDM model, in which the metric of spacetime is Riemannian,
the metric in the TL model is Euclidean, where the angular size 
of an object is inversely proportional to its distance from the observer.
Thus, the predictions of this model are essentially different from
those based on the expanding-Universe concept. 

Consequently, the relationship between the angular $\theta$ and linear $\delta$ sizes 
in TL is 
\begin{equation}
\theta(\delta,z)^{\rm TL} = \frac{H_0}{c} \frac{\delta}{\ln(1+z)}\,, 
\label{eq:tl_ang_size}
\end{equation} 
with the corresponding angular-diameter distance being 
\begin{equation}
D_A^{\rm TL}(z) = \frac{c}{H_0} \ln(1+z)\,. 
\label{eq:tl_and_diam_dist}
\end{equation}

Even before Zwicky's works, static cosmological models were proposed, in which
the cosmological redshift was explained by general-relativistic (i.e., geometrical or
gravitational) change of photon energy with distance. 
The first static general-relativistic cosmological model was introduced
by  \cite{einstein17}. At that time, he was not concerned with the 
cosmological redshift problem because there was not then available 
observational evidence for such a phenomenon, and the Universe was commonly  
believed to be static. 
However, based on Einstein's theory alone, it was already possible
to foresee the existence of the cosmological redshift. 

This was done by W. de Sitter in his prophetic 1917-paper \citep{desitter17c}, where he 
considered positively curved 3-manifolds of spherical, $\mathbb S^3$,  
and elliptical shapes, the latter being also called projective space, $\mathbb P^3$.
The elliptic space in de Sitter's considerations models the physical world
by projecting it onto the Euclidean space $\mathbb E^3$.  
The projection corresponds to the coordinate transformation 
\begin{equation}
r = R \tan \chi\,,
\label{eq:projection}
\end{equation}
where $R^{-2}$ is the constant positive curvature of $\mathbb S^3$ 
or $\mathbb P^3$. Thus, it uses other coordinates instead of ($r$, $\psi$, $\theta$).  
Locally, $\mathbb S^3$ and $\mathbb P^3$ are identical to $\mathbb E^3$. However, 
 such quantities as  { velocity} and  { energy} are related to 
different coordinate systems. Hence, they may change when observed 
in one or another reference frame. De Sitter argues that  since 
the time-component of the metric
$g_{44}= \cos^2 \chi$ diminishes in the elliptical space with the increase of the distance 
parameter $\chi$, then {``{the frequency of light vibrations
diminishes with increasing distance from the origin of coordinates.
The lines in the spectra of very distant stars or nebulae must therefore be 
systematically displaced  towards the red.}''}  

This 1917-prediction (sometimes called the {{de Sitter effect}}) 
that the spectra of remote objects are redshifted in a static 
universe endowed with Einstein's curvature 
was made by de Sitter in the same year when Slipher discovered that the spectra of the majority
of galaxies (84\%), which he was observing were redshifted \citep{slipher17}.
This prediction was made well before 
Lundmark's discovery that Slipher's redshifts of galaxies were proportional to 
their distances \citep{lundmark24}. Furthermore, the Hubble--Lema\^{i}tre law 
\citep{lemaitre27,hubble29} was discovered much later in 1927--1929. 
It is interesting to note that a few years before the Lema\^{i}tre--Hubble's discovery,
Eddington warned against 
the possible wrong interpretation of galaxy redshifts as due to  
their recession. In 1923 he wrote: {{``in de Sitter's theory,
there is the general displacement of spectral lines to the red in distant objects due to 
the slowing down of atomic vibrations which would be erroneously
interpreted as a motion of recession''}} \citep{eddington23}. 
That is why Hubble, in his discovery paper \citep{hubble29}, recites 
the de Sitter effect   as one of the possible mechanisms
responsible for the distance-redshift relationship.
This redshift mechanism in the Einstein--de Sitter's 
model of the Universe was later discussed in more detail by  \cite{hoyle75}.

Based on his prediction, de Sitter estimated the angular size $\theta$ of 
a remote object at a distance $r=R\chi$ from the observer, which 
was one of the first attempts to link the size of the Universe and
cosmological redshift. 

De Sitter's ideas were further developed in 1974 by I.E. Segal in his 
  {Chronometric Cosmology} theory \cite{segal74}.  
He pointed out \citep{segal95} 
that  ``{time and its conjugate variable, energy,
in the Universe with the Einstein curvature 
are fundamentally different  from the conventional time and energy in the local 
flat Minkowski space that approximates the Einstein Universe at the point of observation}''. 
Despite being very closely related to the Einstein--de Sitter model of the Universe,  
Segal's cosmology is rejected by the astronomical community because it fails 
to match observational data. Although he used the correct approach 
by making a distinction between space and time approximations in the curved and flat spacetimes,
Segal arrived at incorrect expressions for the redshift-to-distance law and other 
quantities needed for testing cosmological models.

More recently, de Sitter's idea of using the Einstein Universe curvature 
has been revived by  \cite{marr22} who literally follows the 
logic of the de Sitter's 1917 work, but strangely does not mention it. Nevertheless, 
by using his visualisation tool for representing photon paths in the form of logarithmic
spirals, $1+z =e^\psi$, Marr   derived the following expression for the 
angular-diameter distance  
\begin{equation}
D_A^{\rm Marr}(z) = \frac{c}{H_0} \ln(1+z)\,, 
\label{eq:marr_and_diam_dist}
\end{equation} 
which coincides exactly with the expression 
(\ref{eq:tl_and_diam_dist}) of the TL model due to the use of the exponential 
form of the logarithmic curves representing 
photon paths. 
The Hubble diagram built by Marr with the use
of the luminosity distance based on Equation (\ref{eq:marr_and_diam_dist}) agrees satisfactorily 
with the distance moduli of the type-Ia supernova, having  the same level of accuracy 
as the $\Lambda$CDM-based Hubble diagram or even better.

Despite the obvious possibility of Einstein's curvature being the basis of one of the possible
explanations of the observed cosmological redshift, Einstein himself 
abandoned his static-Universe solution as being  unstable. 
Instead, in 1931, he proposed a cyclic-Universe model \citep{einstein31}; thus,   
siding with the Lema\^{i}tre's dynamical interpretation of the cosmological
redshift. Unfortunately, Einstein did not know that 39 years later his static solution 
would be proven stable by his collaborator \cite{rosen70}.
Nevertheless, Einstein still had his reservations with respect to the 
expanding-Universe model, expressing some doubts in 
his 1931 paper. Commenting on the 
estimated time from the beginning of the expanding Universe he wrote:
  ``{The greatest difficulty of this whole approach is
that the elapsed time since P = 0 comes out at only about $10^{10}$ years}''.
He chose the word ``difficulty'' because he was well aware of the fact that 
this estimated timespan was smaller than the 
ages of some stars, which were found to be of about $10^{13}$ years \cite{condon25}.
A more recent example of a star with an age $14.46 \pm 0.31 Gyr$, which 
exceeds the presumed time from the beginning of the Universe, is HD 140283 \cite{bond13}.

At low redshifts ($z<0.1$), the distance-to-redshift relationship is approximately linear:
\begin{equation}
z = \frac{H_0}{c}r\,, 
\label{eq:tl_zlin}
\end{equation} 
with the corresponding angular-diameter function
\begin{equation}
\theta(\delta,z)^{H_0} = \frac{H_0}{c} \frac{\delta}{z}. 
\label{eq:tl_thetalin}
\end{equation} 

At higher redshifts, different cosmological models deviate from this linear relatioship in diverse ways, 
which allows the use of the angular size measurements to differentiate between 
cosmological models. Here, for  comparison purposes, we shall use 
this simplest linear relationship in our plot of galaxy angular sizes  { (see Section 3)}
to highlight that neither $\Lambda$CDM, nor the linear function, match the JWST data.

\begin{figure}[h]
%
\hspace{1cm}
\includegraphics[width=8cm]{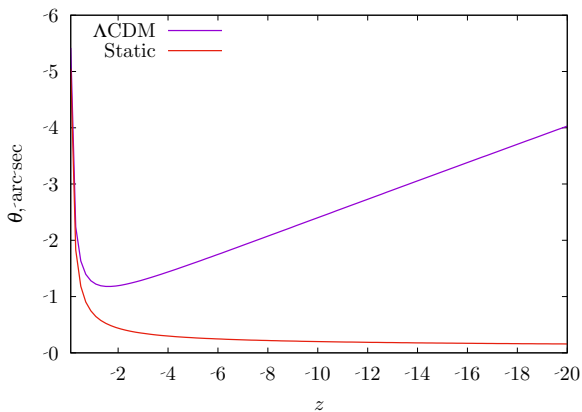} 
\caption{Angular size of a 10 kpc-size object 
 as a function of redshift $z$ within the framework of the static-Universe model 
(the red curve) as compared to the same relationship within the framework of the 
$\Lambda$CDM model (purple curve) for $H_0=70$\,km\,s$^{-1}$\,Mpc$^{-1}$.}
\label{fig2a}
\end{figure}

If we compare the angular diameter functions for the expanding-  and
static-Universe models (see Figure \ref{fig2a}),  
we note that the static (TL) model predicts much smaller angular sizes of high-redshift 
galaxies than $\Lambda$CDM. Therefore, according to this prediction, JWST should observe small 
(by their angular size) galaxies with large surface brightnesses. 
Within the framework of the expanding-Universe model,
a typical 10-kpc-galaxy, as seen from the distance corresponding to $z=14-16$,
would appear as a 3''-angular-size object. Whereas, according to the static-Universe 
model, JWST should observe it to be very small---a fraction of an arcsecond. 
With its large aperture, JWST has a high angular resolution (better that 
0.1''). Thus, it will definitely observe very small galaxies as extended sources
(see the next section).  

\section{Results}

\begin{figure}[h]
%
\hspace{1cm}
\centering
\includegraphics[width=15cm]{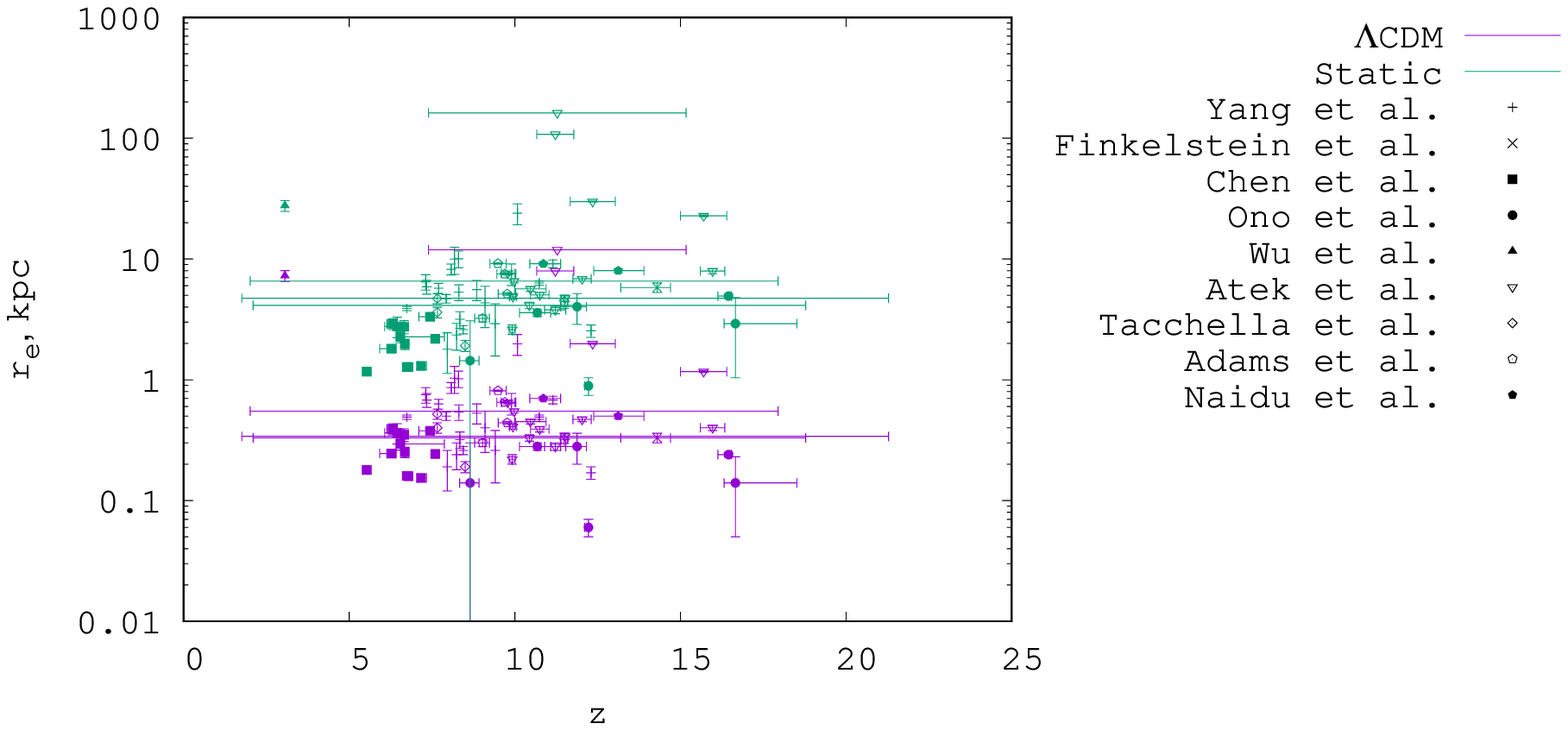} 
\vspace{0.8cm}
\caption{ 
JWST galaxy physical sizes (effective radii $r_e$, in kpc) as estimated within the framework of the 
standard $\Lambda$CDM cosmological model \citep{yang22,finkelstein22,chen22,ono22,wu22,atek22,tacchella22,naidu22a,naidu22b,adams22}
 (purple points) 
and sizes of the same galaxies as they would appear in a static Universe (green points), 
for which we have used the formalism of Zwicky's dissipative (tired-light) model. 
}
\label{fig_size_lcdm_tl_jwst}
\end{figure}

Figure \ref{fig_size_lcdm_tl_jwst} summarises graphically  the 
findings of various research groups that are processing the 
JWST first deep-field images. We are interested here in the correspondence 
between the galaxy sizes (effective radii, $r_e$) and galaxy redshifts. 
The purple points in Figure \ref{fig_size_lcdm_tl_jwst} indicate 
galaxy physical sizes (in kpc) as determined by different authors 
from the observed angular sizes within the standard $\Lambda$CDM 
model. Most of the galaxies found within the JWST field
of view are extremely small, their effective radii varying from 
0.1 kpc to 3 kpc at redshifts $z=6-10$. Assuming their masses are comparable to the 
masses of the local-Universe galaxies ($10^8$ to $10^{11}$\,M$_\odot$),
these galaxies look extremely odd. They have well-developed disks and bulges and
contain dust. Furthermore, their chemical composition is similar to that of local galaxies.
Some of them are likely to contain the same number of stars as the Milky Way,
but they look like a Milky Way squeezed to 1/10th of its size.

\vspace{-0.2cm}
By contrast, if we look at them from the point of view of an observer
in a non-expanding universe (the green points in Figure \ref{fig_size_lcdm_tl_jwst})
we would find that their sizes are comparable to the sizes of local galaxies
($r_e= 2$ to 30 kpc), and the peculiarity of a ``squeezed Milky Way'' disappears. 

\begin{figure}[h]
%
\hspace{1cm}
\includegraphics[width=8cm]{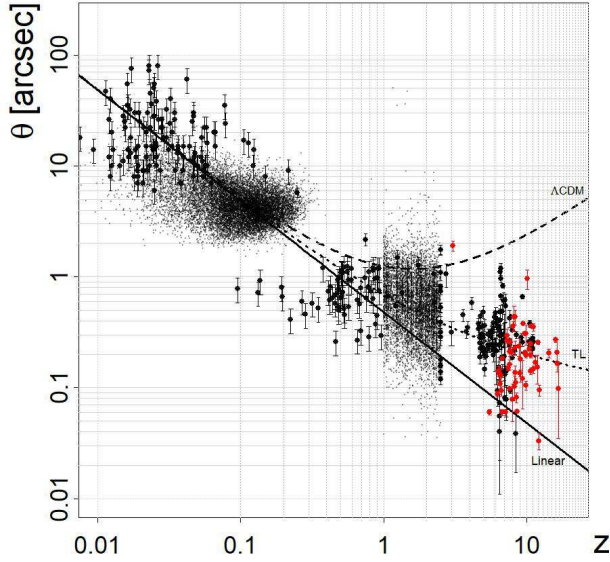}
%
\caption{ 
Angular diameters of a 10-kpc-size object as expected to be seen at different redshifts 
within the frameworks of $\Lambda$CDM  
(dashed curve) and of the non-expanding Universe model, TL (dotted curve).  These 
expectations are compared with the actual angular sizes found 
in the recent JWST observations (red points) and some pre-JWST observations (black points).
The solid curve indicates the simplest linear function for angular diameters
based on the Hubble constant $H_0$. 
}
\label{fig_size_lcdm_tl_jwst2}
\end{figure}

The observed galaxy sizes shown in Figure \ref{fig_size_lcdm_tl_jwst}
are   {also presented (in the form of angular diameters, $2r_e$) } 
in Figure \ref{fig_size_lcdm_tl_jwst2} as red points  {. They}
are compared with the plotted theoretical angular-diameter curves for a 
$\sim 10$\,kpc-object. This is the size of a typical  galaxy in the local Universe 
as it would be seen from distances corresponding to the redshifts $0.1 < z < 20$. 
The scatter of the points in this plot is quite large, and  most of the redshift estimations 
here are photometric, with large error-bars (as we have already mentioned in Section \ref{sec2.1.}, the 
photometric redshifts are within the 
redshift error-bars in Figure \ref{fig_size_lcdm_tl_jwst2} to avoid confusion.    

The theoretical
angular sizes of such objects are plotted for three cases: 
\begin{enumerate}
\item
 An expanding Universe (dashed curve) with 
the expansion parameters corresponding to the standard $\Lambda$CDM cosmological model,
Equation (\ref{eq:dist_angdiam});
\item
  A static Universe with the photon-energy loss and the angular size parametrised according to 
Equations (\ref{eq:tl_and_diam_dist}) or (\ref{eq:marr_and_diam_dist}), dotted curve; 
\item 
 A static Universe with the linear form of the $\theta - z$ relationship,
Equation (\ref{eq:tl_thetalin}), solid curve.
\end{enumerate}

The JWST observations are supplemented in this Figure with some pre-JWST 
observations (black points) made with the use of the Hubble Space
Telescope and some large ground-based telescopes 
\citep{salzer89,koo94,phillips97,zirm07,hathi08,vanderwel14,bowler17,bridge19,bagley22,zavala22}.
In order to get a more definitive result 
we have added to this plot the angular sizes of galaxies from two large
galaxy surveys (small black points), one containing 7003 objects \cite{suess19} with
redshifts from $z=1$ to $\sim 2.5$  and another 
containing 670,722 objects \cite{huchra12}
with redshifts $0 < z < 0.3$. Only 2\% of this latter sample is shown on the plot,
otherwise its statistical properties would be visually  obscured.

\section{Discussion}

The angular sizes of galaxies seen in Figure \ref{fig_size_lcdm_tl_jwst2} 
 exhibit a clear evolutionary trend,
with galaxy angular sizes diminishing to $\theta \approx 0.2"$
as the redshifts increase to $z \approx 15$. 
This is currently understood in terms of their time-evolution,
from some protogalaxies at very high redshifts to large size 
galaxies appearing via merging of smaller galaxies.
There is a  problem here: those small size galaxies detected by the JWST 
at high redshifts are too massive to be protogalaxy candidates. 
This suggests that what is occurring is a pseudo-evolution.  

This (pseudo-)evolution has been noticed by many researches. 
For example,  \cite{ono22}
demonstrate  that  
the effective radii $r_e$ of galaxies tend to diminish from 50 kpc to 0.1 kpc as their redshifts 
increase from $z=0$ to $z = 17$. These authors point out that a few galaxy candidates,
such as GL-z12-1 ($z \approx 12$),  whose sizes are exceptionally small 
($r_e^{\rm GL-z12-1} = 61 \pm 11$\,pc), have masses  too high
($\sim 3\times 10^8$\,M$_\odot$) for them to be protogalaxies. 

Other researches \citep{naidu22a} have found a similar 
compact galaxy at  $z \approx 17$, and its mass is
also very big: $M_* \approx 5 \times 10^9$\,M$_\odot$. 
According to the $\Lambda$CDM-approach, 
this galaxy was formed a mere $\sim 220$\,Myr after the Big Bang. 
Numerical simulations  predict galaxy
masses to be below $10^8$\,M$_\odot$ at this cosmic time \citep{pallottini22}.
So, the authors of \citep{naidu22a} came to the conclusion
that such a galaxy  ``{challenges virtually every early galaxy
evolution model that assumes $\Lambda$CDM cosmology}''.
They also point out that, given the relatively small area currently explored
by JWST (less than 60 square arcminutes), the number of very bright objects found
within this area is surprisingly large. For example, \cite{naidu22b} have found two very bright galaxy candidates 
at $z=11$ and $z=13$ and calculated the corresponding number densities (UV-luminosity functions)
$\phi_{UV} \approx 6.3\cdot 10^{-7}$\,[mag$^{-1}$ Mpc$^{-3}$] for $M_{UV} \approx -22.1$
and  $\phi_{UV} \approx 1\cdot 10^{-5}$\,[mag$^{-1}$ Mpc$^{-3}$]  for $M_{UV} \approx -20.8$.
If we compare these values with similar estimations made by Bowler et al. \cite{bowler20}  
for lower redshifts ($z=9$),
$\phi_{UV} \approx 8\cdot 10^{-7}$\,[mag$^{-1}$ Mpc$^{-3}$]  for $M_{UV} \approx -22.1$
and $\phi_{UV} \approx 9\cdot 10^{-5}$\,[mag$^{-1}$ Mpc$^{-3}$]  for $M_{UV} \approx -20.8$,  
then we see that number densities of bright galaxies are practically the same for 
$9 <z <13$, although theoretically they should substantially diminish at higher redshifts. 

This discrepancy also follows from the results of hydrodynamic simulations of
the Millen\-niumTNG project \citep{kannan22}. It turns out that 
beyond $z \ge 12$, this simulation underpredicts the abundance of luminous galaxies 
and their star-formation rates by almost an order of magnitude. 
The authors of this simulation comment that the same discrepancy is typical for 
most other similar works. They suggest an explanation that there might be some missing 
physical processes that are not included in simulations. However, as we shall see 
in Figure \ref{fig_jwst_masses} below, this discrepancy might be simply due to the
underestimation of cosmic time.   

Standard UV-luminosity functions predict 
a much smaller number of bright objects within 60 square arcminutes.
That is why the authors of \citep{naidu22a}  hypothesise that  JWST is discovering new,
hitherto unknown galaxy populations,   { which was previously suggested by \cite{disney12} 
for explaining HST observations} (alas, Occam's razor
{\footnote{\it{{Pluralitas non est ponenda sine necessitate}}   {\rm(William of Occam)}}}  
is yet again abandoned or forgotten here).

As for the galaxy mergers, as far as we know, the early JWST observations have revealed 
only one candidate of a merging galaxy pair  MACS0647–JD (possibly a triplet) at $z \approx 11$ \citep{hsiao22}.
The theoretical merger rate for $z > 6$ is estimated to be $\sim 0.1$\,[Gyr$^{-1}$] per galaxy 
\citep{ventou17}.
Since JWST has detected about a dozen  galaxy candidates at $z \sim 11$,
the theoretically expected number of detected galaxy mergers approximately matches the observed number, given the cosmic 
time $\sim 0.43$\,Gyr corresponding to $z \sim 11$. However, this theoretical merger rate is likely to be 
underestimated because it does not differ from the merger rate estimated for low-redshift galaxies, whereas the current galaxy 
formation theory expects small galaxies at high redshifts to form large galaxies at low redshift by multiple mergers.
That is why other simulations of galaxy formation give much higher expectation values for the 
theoretical merger rate, $\sim 4-5$\,[Gyr$^{-1}$] per galaxy for $z=10$ \citep{rodriguez15}.  

\begin{figure}[h]
%
\hspace{0.8cm}
\begin{tabular}{ll}
\includegraphics[width=8cm]{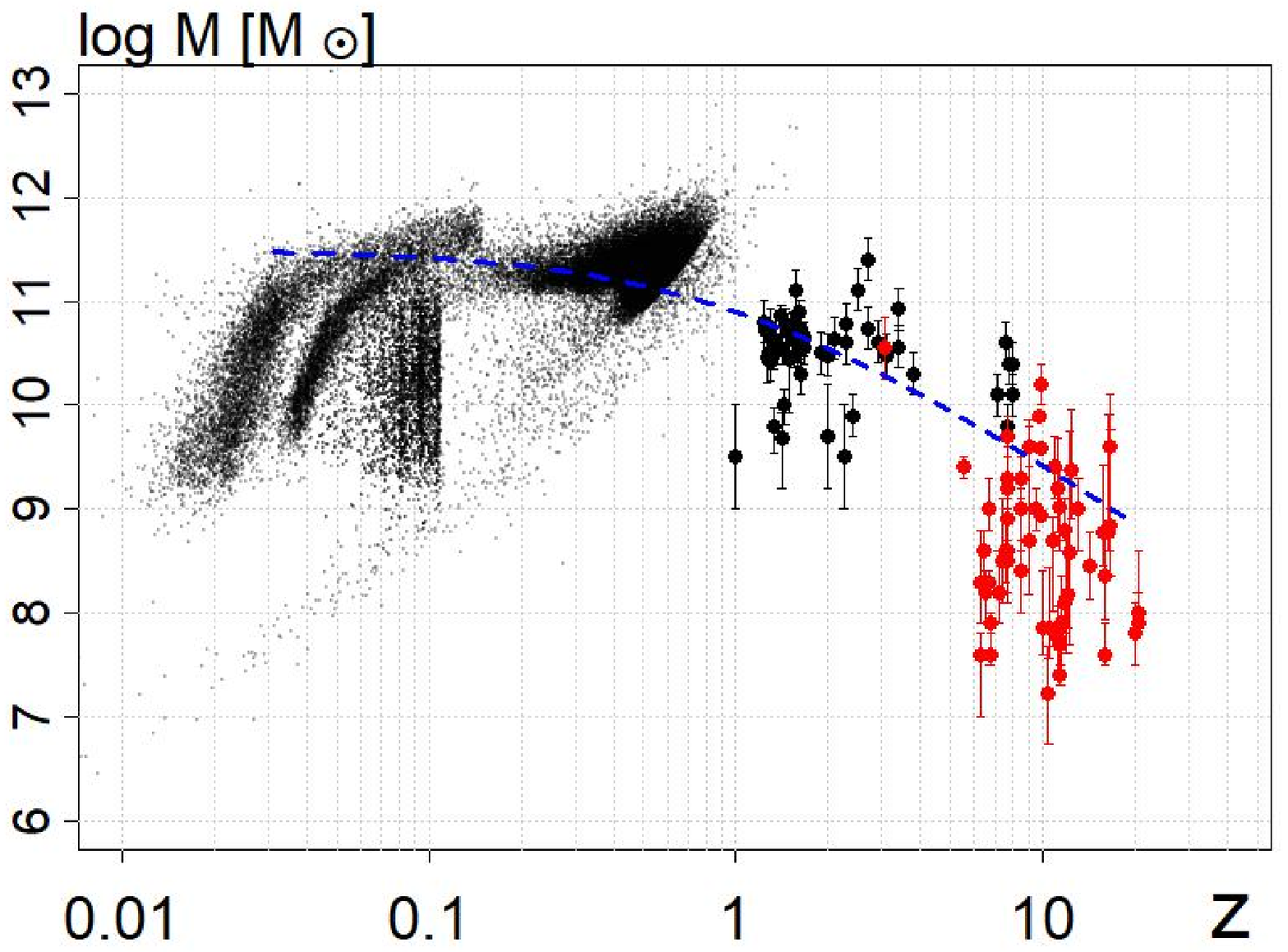} &
\includegraphics[width=8cm]{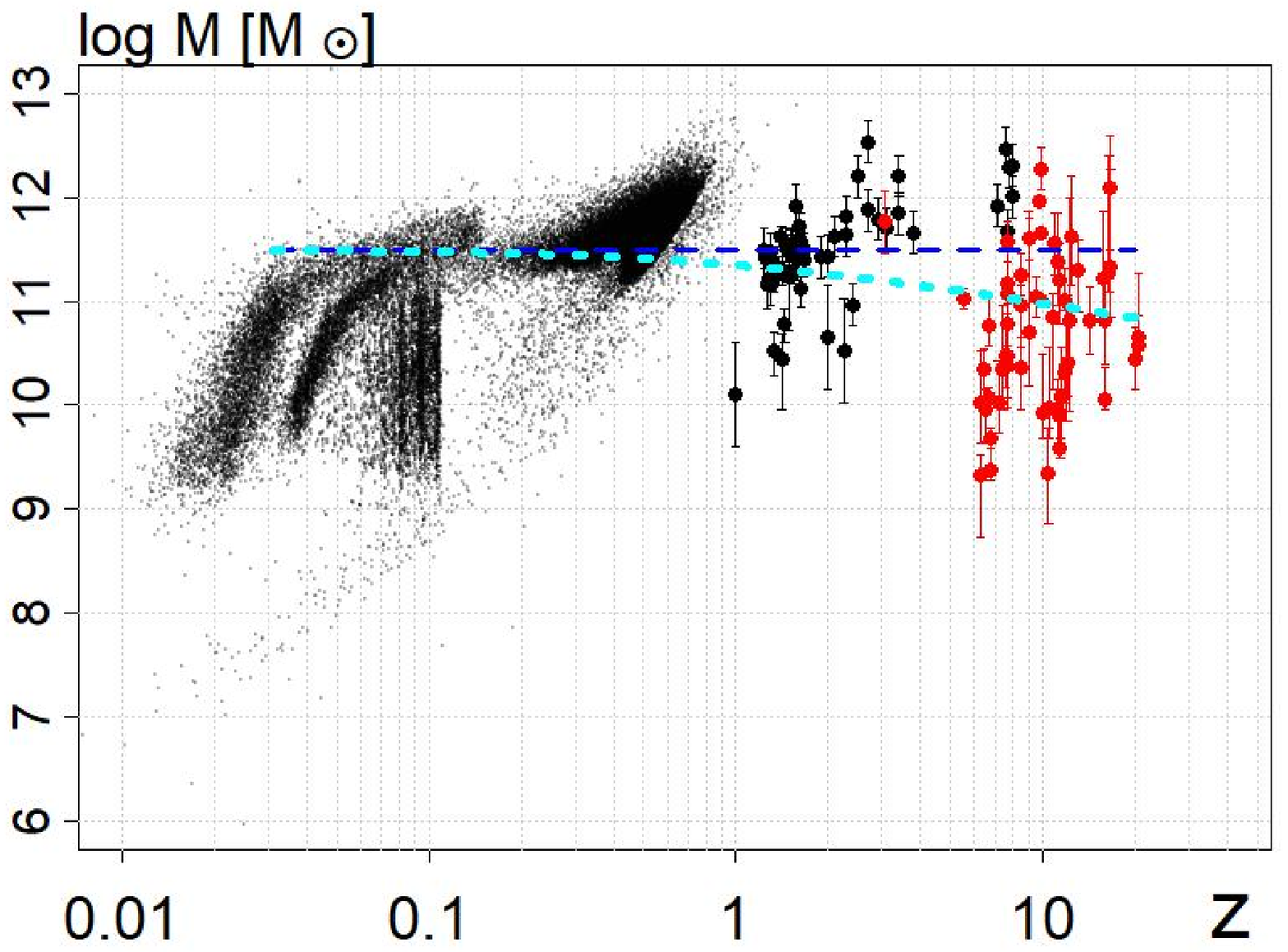}
\\       
\end{tabular}
\caption{ 
   {Left}: Masses of high-redshift galaxies as determined within 
the framework of $\Lambda$CDM  using the recent JWST observations (red points) 
and some pre-JWST observations (black points); the dashed curve
indicates the factor $(1+z)^{-2}$ of the distance-luminosity 
relationship in the standard cosmology;
    {Right}: the same masses
corrected for the factor  $(1+z)^{-2}$ in order to transform them to
 the static-Universe framework.}
\label{fig_jwst_masses}
\end{figure}

The masses of the high-redshift galaxies estimated by using JWST 
observations \citep{naidu22a,naidu22b,yan22,suess22,tacchella22,wu22,ono22,adams22,atek22,chen22,finkelstein22} 
are shown in Figure \ref{fig_jwst_masses} (red points).
The black points indicate masses deduced from some pre-JWST
observations \citep{bridge19,ding20,suess19,calvi11,maraston13,sanchez16} 
made by using large ground-based telescopes and the Hubble Space Telescope.

We note the evolutionary (pseudo-evolutionary) decline in galaxy masses 
toward the high-redshifts (which is highlighted by the dashed curve on the 
left panel of Figure \ref{fig_jwst_masses}). If real, this evolution feeds 
into the $\Lambda$CDM narrative.
Although, there are still some problems related to the lack of time for the possible build-up
of protogalaxies.  
It is difficult to ignore the fact that these alleged protogalaxies are
fully developed, smooth (i.e., they are not disturbed by merging with other galaxies), 
with their disks, bulges and a chemical composition similar
to the chemical composition of local galaxies.\footnote{actually, it is not completely 
ignored by astrophysicists, and the majority of them
are thinking about (contriving of) new possibilities in order to theoretically 
accelerate the process of galaxy formation immediately after the Big Bang, by 
introducing, for example, non-trivial non-Gaussianities into the initial conditions of 
the cosmological perturbations  \citep{biagetti22}, contrary to Occam's principle. 
While some others embrace the idea that the Universe might be much older than what
follows from the $\Lambda$CDM theory and publish their arguments \citep{subramani19} 
or report this idea to the general public via 
documentaries produced by influential media like the BBC 
\url{https://www.youtube.com/watch?v=vAxgaTvYA7Y} (accessed on 1 October 2022).}
It is clearly pointing out that these galaxies are practically the same
as our nearby galaxies in the late Universe.

Concerning this evolutionary trend, we have plotted it in the left-hand side panel 
of Figure \ref{fig_jwst_masses} in the form of 
the function $(1+z)^{-2}$ (the dashed curve). We know that galaxy masses are estimated from 
their luminosities (see, e.g., \cite{mihalas68,freeman70}), and the luminosity--distance function in the 
expanding-Universe models is reduced by the same factor of $(1+z)^{-2}$. 
If we correct the galaxy masses estimated within the $\Lambda$CDM framework
by this factor, we see that not only morphologies and chemical composition 
of the remote galaxies are similar to the local-Universe's galaxies 
  {but their masses as well} (see the right panel of Figure \ref{fig_jwst_masses}). 

Still there is some  noticeable evolutionary trend remaining 
in the high-redshift galaxy masses (indicated by the dotted light-blue 
curve in the right-hand side panel of Figure \ref{fig_jwst_masses}). This means that remote 
galaxies, indeed, grow and evolve, but this happens at a much slower pace 
than is assumed by the standard cosmological model. 

It must also be noted that the cosmological test based on the redshift-dependent 
angular-sizes of galaxies alone do not provide sufficiently strong evidence 
against the $\Lambda$CDM model since the observed evolution of galaxy sizes 
might be real.  Therefore, it would be important to perform other cosmological tests,
by studying, for example, the relationship between the redshift and the 
number-density of high-redshift galaxies and other objects, such as galaxy clusters or quasars. 
In static- and expanding-Universe models, this relationship is different, 
the distinction is of the factor $(1+z)^3$, which is quite large.
 
Within the $\Lambda$CDM model, the number-density of remote objects within the 
spherical layers of thickness $\Delta z$ is fixed (by definition). However, the volumes
of these spherical layers strongly decrease with the increase of $z$. 
This should lead to dramatic increases of the proper (metrical) number-density 
of objects in the high-redshift volumes. Eventually, this number-density 
would become an impossible quantity from the point of view of any physical model. 
The same would be (and is) seen in the number-density of stars 
within the volumes of high-redshift galaxies, as their physical sizes are 
strongly decreased when estimated within the expanding-Universe formalism, 
while the stellar masses of these galaxies remain approximately the same as  the local-Universe galaxies.   

Another important distinctive parameter is the     
cosmological time dilation. It can be used for determining the nature 
of the cosmological redshift.  For example, the 
static cosmological model with photon-energy dissipation predicts no 
time dilation. Whereas the FRLW models and the static models with the 
cosmological redshift of the general-relativistic (gravitational) nature
predict the time-dilation factor 
scaling with the factor of $(1+z)$.  

The time-dilation effect is actually observed in the light-curves of the type-Ia 
supernovae \citep{leibundgut96,guy07,blondin08}, 
which strongly supports the expanding-Universe models or the 
static-Universe models of the Einstein-de-Sitter type.   
Nevertheless, additional studies of this effect are still needed because
there exist some evidence against the cosmological time-dilation.  
In particular, gamma-ray bursts, nova-like stars,
quasars and fast radio-bursts are poorly explored in terms of their light-curve 
duration. Available publications with the results of temporal behaviour 
of their light-curves in relation to their redshifts are contradictory
\citep{hawkins10,horvath22}.
New research in this direction would reduce systematic errors
and check more rigorously the functional dependence of the time-dilation
factor on $z$. It would be important for revealing possible 
combined types of cosmologies (e.g., the TL-redshift mixed with the
the redshift due to the Universe  expansion). 

Further study of galaxy angular-sizes and number-densities at high reshifts is also very important 
for determining more accurately the cosmological-model parameters. 
Additionally, finding transient objects at high-redshifts,such as supernovae, gamma-ray bursts 
or fast radio-bursts, and measuring parameters of their light-curves 
would help specify more accurately the nature of the cosmological time-dilation effect. 
New goal-settings would likely emerge on the course of these studies, as there
might be some unexpected findings on this way.

Consequently, it would be very important to continue gathering and analysing JWST observations
of the high-redshift Universe, compiling large catalogues 
of photometric and spectrophotometric redshifts of remote objects. 

Although our analysis of JWST observations favours a static (TL) model 
of the Universe rather than the standard (expanding-Universe) cosmology, the latter is strongly supported 
by other observational evidence. The main challenges which any static cosmological model must
face are  the cosmic microwave background (CMB), the 
abundance of chemical elements in the Universe and the formation of cosmic structure. 

In the standard model, the CMB, with its black-body spectral energy distribution with $T=7$\,K 
was predicted by  \cite{gamow53} to exist before it was actually discovered. 
It would be fair to acknowledge that for the static Universe, a similar thermalised 
radiation with $T=3$\,K was predicted even earlier by   \cite{eddington26}
and  with $T=2.8$\,K by  \cite{nernst37}. For more comprehensive reviews
as to the possibility of explaining the CMB radiation within the framework of  static-Universe models
see, for example, \cite{baryshev96,cirkovic18}.

There are also static-Universe alternatives to the standard model predictions with respect to
the light element abundances and baryon fraction. For example, 
G.R. Burbidge and F. Hoyle  discussed the possibility of helium creation in massive objects 
\citep{burbidge71, burbidge98}. \cite{salvaterra03} proposed that
the primordial He abundance could be created by population III stars. 
On the other hand, the standard Big Bang nucleosynthesis theory is not without serious problems
\citep{sargent67, terlevich92, izotov10}.

The only open issue for static-Universe models that remains is the origin of the overdensities, which
leads to cosmic structure formation and which is elegantly solved in the standard cosmology 
by the mechanism of initial quantum fluctuations and baryonic acoustic oscillations. 
However, although we have used here the tired-light model as a static-Universe model example, 
from the right-hand side panel of Figure \ref{fig_jwst_masses} we see that the Universe is
evolving with respect to this model. That means the real picture is more complicated than a 
pure static cosmological solution. Therefore, the physical mechanism for the original overdensities and 
cosmic structure formation could well be the same as in the $\Lambda$CDM cosmology, 
including quantum fluctuations and the baryonic acoustic oscillations.
Although with respect to the latter, various authors mention that there might be 
some problems with their statistical analysis and accuracy 
\citep{baryshev12,gabrieli05,nabokov10,shirokov16}.

\section{Conclusions}
We conclude that the first JWST observations of high-redshift objects
cannot be explained by the expanding-Universe model. 
Everything points to the possibility that 
the actual age of the objects in the Universe is far larger than predicted by 
$\Lambda$CDM cosmology. Of course, we should be cautious about such a conclusion.
Thus, before dismissing the expanding-Universe paradigm, it is important to 
robustly confirm the new findings. 

No doubt, much longer exposures and much 
deeper fields will be acquired in the forthcoming years by the JWST. 
These longer exposures would likely result in new galaxies discovered 
at $z \approx 20$ or more. Based on our conclusion, we predict that
the JWST should discover even smaller galaxies 
(in terms of their angular-sizes) and that those smaller galaxies would be observed
as very luminous, with well-developed morphology.
They would be approximately the same (perhaps, slightly less-evolved) as the 
galaxies of the late Universe. In such a case, the expanding-Universe paradigm 
would require correction and modification, in line with the discussion
presented here.

\section*{Data availability}
This work is based on observations made with the NASA/ESA
Hubble Space Telescope (HST) and NASA/ESA/CSA James Webb
Space Telescope (JWST) obtained from the Mikulski Archive
for Space Telescopes (MAST) at the Space Telescope Science
Institute (STScI).
The data are publicly available  
at \url{https://archive.stsci.edu}  (accessed on 1 October 2022), which is the interface to the 
HST and JWST data provided by the 
{\tt Mikulski Archive for Space Telescopes} (MAST) 
of the Space Telescope Science Institute (STScI) under program ID 2736.
The associated programmatic interface 
\url{https://astroquery.readthedocs.io/en/latest/mast/mast.html}  (accessed on 1 October 2022)
provides scripts for data access and reduction. 
Some results of the JWST data reduction are also made publicly available. For example,
the calibrated and distortion-corrected NIRCam and NIRISS images processed by G. Brammer
are accessible via  \url{https://s3.amazonaws.com/grizli-v2/SMACS0723/Test/image_index.html}  (accessed on 1 October 2022). 
The catalogues of ultra-high-redshift objects detected by the JWST instruments in the 
SMACS-0723 deep field are publicly available at 
\url{https://zenodo.org/record/6874301\#.YubQUfHMJes} (accessed on 1 October 2022).

\section*{Acknowledgements}
{We would like to thank Dr. Alice Breeveld
and Dr. Leslie Morrison for useful discussions on the matters in this
paper. We are also grateful to two anonymous reviewers for their comments and very useful suggestions for improving our manuscript.}

\section*{Conflict of interests}
{The authors declare no conflict of interest.}


\end{document}